\documentclass[aps,prd,floats,amssymb,showpacs,nofootinbib,preprint]{revtex4}
\usepackage{graphicx}
\usepackage{amsmath}
\usepackage{amssymb}
\usepackage{epsfig}

\newcommand{\be}{\begin{equation}}
\newcommand{\ee}{\end{equation}}
\newcommand{\bea}{\begin{eqnarray}}
\newcommand{\eea}{\end{eqnarray}}

\begin{document}


\title{Can Cosmic Parallax Distinguish Between Anisotropic Cosmologies?}
\author{Michele Fontanini$^{a,b}\!\!$ \footnote{mfontani@physics.syr.edu}}
\author{Mark Trodden$^{a}\!\!$ \footnote{trodden@physics.upenn.edu}}
\author{Eric J. West$^{a,b}\!\!$ \footnote{ejwest@physics.syr.edu}}

\affiliation{
$^a$Center for Particle Cosmology, Department of Physics and Astronomy, University of Pennsylvania,
Philadelphia PA 19104, USA \\
$^b$Department of Physics, Syracuse University, Syracuse NY 13244, USA}

\date{\today}

\begin{abstract}

In an anisotropic universe, observers not positioned at a point of special symmetry should observe {\it cosmic parallax} - the relative angular motion of test galaxies over cosmic time. It was recently argued that the non-observance of this effect in upcoming precision astrometry missions such as Gaia may be used to place strong bounds on the position of off-center observers in a void-model universe described by the Lemaitre-Tolman-Bondi metric. We consider the analogous effect in anisotropic cosmological models described by an axisymmetric homogeneous Bianchi type I metric and discuss whether any observation of cosmic parallax would distinguish between different anisotropic evolutions.

\end{abstract}

\maketitle


\section{Introduction}

Homogeneity and isotropy are the cornerstones of the standard cosmological model, providing not only a tremendous simplification of General Relativity, but remarkable agreement with all observations. Homogeneity is supported by the observed galaxy distribution from large-scale structure surveys, while isotropy is supported by, in particular, the remarkable uniformity of the average temperature of the cosmic microwave background radiation (CMB). Nevertheless, the paradigm-changing observation of cosmic acceleration, now more than a decade old, has forced cosmologists to re-examine even their most cherished assumptions, including the correctness of General Relativity, a vanishing cosmological constant, and, more recently, homogeneity and isotropy.

The primary evidence for the accelerating universe comes from the unexpected dimming of type Ia supernovae~\cite{Riess:1998cb, Perlmutter:1998np, Bahcall:1999xn, Peebles:2002gy, Copeland:2006wr, Frieman:2008sn, Caldwell:2009ix, Silvestri:2009hh}, as measured through their light curves.  The connection to cosmic acceleration requires the assumptions of homogeneity and isotropy, and thus this raises the possibility that abandoning one of these principles may allow for the appearance of accelerated expansion without actual acceleration itself. 

Of course, the usual cosmological Friedmann-Robertson-Walker (FRW) metric is so simple by virtue of its underlying symmetries. Abandoning these leads to a correspondingly more complicated metric. It is convenient therefore to begin by studying toy models. One class of these that has shown some promise in this direction are the {\it Lemaitre-Tolman-Bondi} (LTB) metrics, in which we are assumed to live inside a spherically symmetric underdense region of spacetime (or ``void'') embedded in an otherwise spatially flat and homogeneous Einstein-de Sitter universe~\cite{Tomita:2000jj, Tomita:2001gh, Tomita:2002td, Moffat:2005yx, Alnes:2005rw, Inoue:2006rd, Inoue:2006fn}. Such a spacetime is manifestly inhomogeneous, due to the void, and on its own violates any strong version of the Copernican principle, since we must live inside this void in order to account for the observed supernovae dimming. Nevertheless, it has been shown that these models can provide a satisfactory fit to the luminosity distance-redshift relation of type Ia supernovae and the position of the first peak in the CMB~\cite{Alnes:2005rw,Chung:2006xh}. Thus, LTB models have been suggested as a possible solution to the problem of cosmic acceleration, obviating the need for quintessence fields, modifications of gravity, or a cosmological constant, and considerable effort has been devoted to constraining them~\cite{Alnes:2006pf, GarciaBellido:2008nz, GarciaBellido:2008gd, GarciaBellido:2008yq}.

Beyond the usual cosmological tests of homogeneity and isotropy, it has recently been suggested that these models could be further constrained by {\it real time} cosmological measurements - in particular, precision measurements of the evolution of the angular positions of distant sources. Following the authors of~\cite{Quercellini:2008ty}, we refer to this effect as {\it cosmic parallax}. The expansion of an FRW universe is isotropic for all observers, and so cosmic parallax would not be observed. Of course although our universe is very close to an FRW universe, it is not exactly so--for example, on large scales bound structures may acquire small peculiar velocities, giving rise to a slight deviation from observed isotropic expansion. However, to any observer living off-center inside the void of an LTB universe, cosmic evolution itself is anisotropic and is an additional source of cosmic parallax. For sufficiently off-center observers the cosmic parallax due to anisotropic expansion would dominate over the contribution from peculiar velocities. Cosmic parallax could therefore provide an interesting test of void models. Upcoming sky surveys such as Gaia \cite{Perryman:2001sp} may be able to initiate a measurement of this effect, requiring only that a similar survey be completed $10$ years later in order to complete the measurement. The absence of cosmic parallax beyond what is expected from peculiar velocities would put an upper bound on how far our galaxy could be from the center of the void in otherwise allowed LTB models. For example, the authors of~\cite{Quercellini:2008ty} argue that Gaia may map sufficiently many quasars, with enough accuracy, so that two such surveys spaced $10$ years apart could detect the additional LTB contribution to cosmic parallax if the Milky Way is more than $10$ megaparsecs from the center of a $1$ Gpc void. If after a decade no additional contribution were found, that would constrain the Milky Way to lie unnaturally close to the center of such a void.

On the other hand, detection of a contribution to the cosmic parallax not arising from peculiar velocities would indicate that the expansion of the universe is anisotropic from our vantage point. In LTB models this would be due to living away from the center of the void, and the strength of the additional contribution would be related to this distance. But cosmic parallax not attributable to peculiar velocities is a generic feature of any cosmological model with anisotropic expansion--an observation also made in~\cite{Quercellini:2008ty}. For example, spacetimes of the Bianchi type would exhibit an additional contribution to cosmic parallax around every point. These homogeneous and anisotropic spacetimes have recently been invoked during inflation to explain anomalies in the CMB~\cite{Ackerman:2007nb,Jaffe:2005pw,Jaffe:2006sq}\footnote{Such theories may not, however, be without problems~\cite{Himmetoglu:2008zp}.} and during late-time cosmology to describe the expansion driven by anisotropic dark energy~\cite{ArmendarizPicon:2004pm,Koivisto:2008ig}. The fact that motivations exist for studying both LTB and Bianchi spacetimes, and that both may exhibit contributions to cosmic parallax that are not attributable to peculiar velocities, raises the question of how we might interpret any additional cosmic parallax signal. Although an observed signal would provide evidence for deviations from an FRW universe, such deviations could in principle be due to deviations from spatial homogeneity (as in LTB models) or deviations from isotropy (as in Bianchi models). It is therefore of interest to consider how cosmic parallax in Bianchi models differs from LTB models.

The structure of the paper is as follows. In section~\ref{ltb} we briefly review the results of~\cite{Quercellini:2008ty}. In section~\ref{bianchi} we describe the kinds of anisotropic models that we will focus on; namely a subclass of Bianchi Type I models. Since we know from observations of the CMB that the universe is very nearly isotropic at the time of last scattering, we describe how we restrict ourselves to Bianchi Type I metrics which pass existing cosmological tests. In section~\ref{parallax} we derive the geodesic equations for these spacetimes and discuss our procedures for numerically integrating them. We then present the cosmic parallax signal we find for these models and discuss how it compares to the cosmic parallax in void models.


\section{Cosmic Parallax in an LTB Void}
\label{ltb}

We begin by briefly reviewing how cosmic parallax manifests itself in LTB models, as discussed in~\cite{Quercellini:2008ty}. Here and throughout the rest of the paper, unless otherwise indicated, by cosmic parallax we mean cosmic parallax due to anisotropic expansion about an observer. The LTB metric is given by~\cite{Lemaitre:1933gd,Tolman:1934za,Bondi:1947av}
\be
 ds^2 = -dt^2 + \frac{|R'(t,r)|^2}{1+\beta(r)}dr^2 + R^2(t,r) d\Omega^2 \ ,
\ee
where $R(t,r)$ is a position-dependent scale factor, $\beta(r)$ is related to the curvature of the spatial slices, and $()'\equiv \partial/\partial r$. The Einstein equations relate $R(t,r)$ to $\beta(r)$ and an additional arbitrary function of integration $\alpha(r)$. Specifying $\alpha(r)$, $\beta(r)$, and an initial condition for $R(t,r)$ completely determines the spacetime. In models with an underdense region, or ``void'', surrounded by an overdense region, $\alpha(r)$ and $\beta(r)$ roughly correspond to the width of the void and the gradient of the boundary between the inner and outer regions, respectively, and will be specified below. 

The LTB metric describes a region of spacetime that is isotropic about the origin but inhomogeneous with respect to the radial direction. Therefore distant galaxies appear to be receding at the same rate in all directions for observers located at the origin. On the other hand, observers located away from the center of the void could in principle observe anisotropic recession. One way to observe this effect~\cite{Quercellini:2008ty} is to measure how the angle between the positions of two distant sources evolves over time. This difference is referred to as the cosmic parallax. To study this, one considers null geodesics in an LTB universe, obeying
\be
   \frac{d^2{}x^\mu}{d\lambda^2} 
   + \Gamma^\mu{}_{\rho\sigma}\frac{dx^\rho}{d\lambda}\frac{dx^\rho}{d\lambda} = 0\ .
   \label{geodesic}
\ee
Here $\Gamma^\mu{}_{\rho\sigma}$ are the Christoffel symbols, $\lambda$ is an affine parameter along null geodesics, and the four-velocities $u^\mu\equiv\frac{dx^\mu}{d\lambda}$ satisfy $u^\mu u_\mu=0$. The goal will be to solve these equations to determine the null geodesics along which light travels from various sources to an observer, and to do this for two different observation times. 

\begin{figure}[h]
   \begin{center}
   \includegraphics[width=0.4\textwidth]{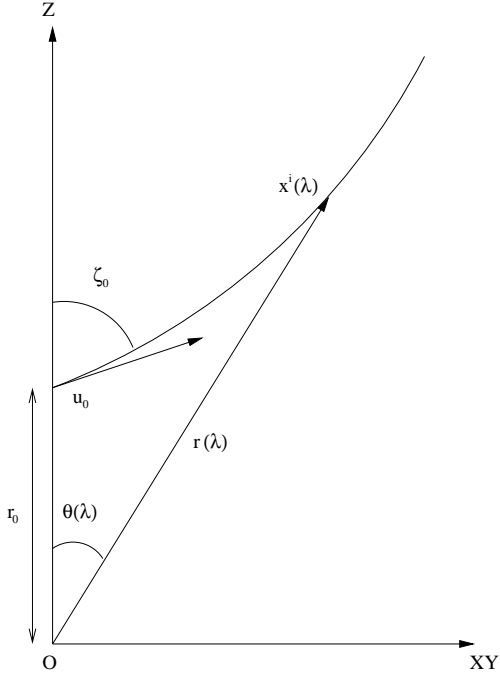}
   \caption{In the LTB model the observer is located at a distance $r_0$ from the center of the void along the axis of symmetry. Each point along the geodesic is described by the spatial coordinates $(\theta(\lambda),r(\lambda))$ or equivalently by $x^i(\lambda)$}
   \label{fig1}
   \end{center}
\end{figure}

Following~\cite{Quercellini:2008ty}, we work in spherical-polar coordinates $(t,r,\theta,\phi)$ for which the origin coincides with the center of the void (labelled ``O'' in figure \ref{fig1}). 
Without loss of generality we choose the observer to lie on the polar axis at a coordinate distance $r_0$ along it. Spherical symmetry about the observer is now broken but the remaining cylindrical symmetry applied to~(\ref{geodesic}) allows us to neglect the polar angle $\phi$ dependence. The system then reduces to three second-order geodesic equations, or equivalently six first-order equations. Applying the null geodesic condition further reduces the system to five independent first-order equations for $t$, $r$, $\theta$, $p\equiv dr/d\lambda$ and the redshift $z$ as~\cite{Quercellini:2008ty,Alnes:2006pf}
\bea
   \frac{dt}{d\lambda} &=& -\sqrt{\frac{(R')^2}{1+\beta}p^2 +\frac{J^2}{R^2}} \\
   \frac{dr}{d\lambda} &=& p \\
   \frac{d\theta}{d\lambda} &=& \frac{J}{R^2} \\
   \frac{dz}{d\lambda} &=& \frac{(1 + z)}{\sqrt{\frac{(R')^2}{1+\beta}\,p^2 +\frac{J^2}{R^2}}}
    \left[\frac{R'\dot{R}'}{1+\beta}p^2 + \frac{\dot{R}}{R^3}J^2\right] \\
   \frac{dp}{d\lambda} &=& 2\dot{R}'p\sqrt{\frac{p^2}{1+\beta} 
   + \left(\frac{J}{R R'}\right)^2}
   \mbox{} + \frac{1+\beta}{R^3 R'}J^2
   + \biggl(\frac{\beta'}{2+2\beta} - \frac{R''}{R'} \biggr) p^2\ ,
\eea
where $J\equiv R^2 d\theta/d\lambda = J_0$, is constant along the geodesic. 

To completely specify the system we require five initial conditions, which we provide for convenience at the initial observation event, and denote with a subscript ``0''. Since one would like to specify initial conditions in terms of physically measurable quantities, we consider the angle $\xi_0$ between the polar axis and the line of sight along an incoming photon trajectory arriving at the observer (see figure~\ref{fig1}).  This angle $\xi_0$ coincides with the coordinate angle $\theta$ when $r_0=0$, but in general it is given by~\cite{Alnes:2006pf}
\be
   \cos{\xi_0} = -\left.\frac{R'(t,r) p}{\frac{dt}{d\lambda} \sqrt{1+\beta(r)}}\right|_{\lambda=0}\ .
\ee
This expression can be used to express $J_0$ and $p_0$ in terms of $t_0$, $r_0$, and $\xi_0$, via 
\bea
   J_0 &=& R(t_0,r_0) \sin{\xi_0} \nonumber \\
   p_0 &=& \frac{\cos{\xi_0}}{R^\prime(t_0,r_0)}\sqrt{1+\beta(r_0)}\ .
\eea
Therefore, the system is completely determined by specifying $t_0$, $r_0$, $\theta_0$, $z_0$, and $\xi_0$. 

Clearly, our coordinate choice means that $\theta_0=0$, and our conventional definition of redshift yields $z_0=0$. Following~\cite{Quercellini:2008ty} we choose $r_0=15 {\rm Mpc}$, which is the largest value consistent with the CMB dipole~\cite{Alnes:2006pf}. What remains is to specify a direction on the sky and the time of observation to uniquely determine a geodesic that terminates at the spacetime event of observation. This picks out an initial geodesic along which light from a distant source travels to reach the observer.

Given the redshift and line of sight angle of a source at an initial time, we can determine the trajectory of light from that source at a later time in the following way. As just mentioned, the line of sight angle $\xi_0$ picks out an initial geodesic, terminating at the initial observation event. The observed redshift determines how far backwards in time along this initial geodesic the source lies. We extract the comoving coordinates of the source by numerically integrating backwards along this initial geodesic. Once the comoving coordinates of the observer and source are determined, and the interval of time between observations is specified, we solve a boundary-value problem to determine the final geodesic. We then extract the line of sight angle of this final geodesic $\xi_f$.

After repeating this procedure for two sources we have four angles: the input angles $\xi_{a0}$ and $\xi_{b0}$ and the output angles $\xi_{af}$ and $\xi_{bf}$. We may then compute the difference 
\be
   \Delta\gamma\equiv \gamma_f-\gamma_0=(\xi_{af}-\xi_{bf})-(\xi_{a0}-\xi_{b0}) \ ,
   \label{cpdef}
\ee
which is the main quantity of interest, hereafter referred to as the cosmic parallax. 

In~\cite{Quercellini:2008ty} this quantity is calculated for LTB models characterized by the functions
\bea
   \beta(r) &=& \left(H_0^{\rm OUT}\right)^2 r^2
   \frac{\Delta\alpha}{2}\left(1 - \tanh \frac{r-r_{\rm vo}}{2\Delta r}\right)
\\
   \alpha(r) &=& \left(H_0^{\rm OUT}\right)^2 r^3 - r\beta(r) \ ,
\eea
which characterize the smooth transition from the inner underdense region to the outer, higher density region. The quantities $\Delta\alpha$, $r_{\rm vo}$, $\Delta r$, and $H_0^{\rm OUT}$ are free parameters of the model that can be tuned to fit CMB and supernovae measurements. Following~\cite{Quercellini:2008ty} and~\cite{Alnes:2005rw},  we choose a model which is in good agreement with SNIa observations and the location of the first peak of the CMB, namely $\Delta\alpha=0.9$, $r_{\rm vo}=1.3 \, {\rm Gpc}$, $\Delta r=0.4 \, r_{\rm vo}$, and $H_0^{\rm OUT}=51 \, {\rm km \, s^{-1} \, Mpc^{-1}}$. Although this Hubble parameter outside the void seems to be in conflict with the measured Hubble Key project value of $72 \pm 8 \, {\rm km \, s^{-1} \, Mpc^{-1}}$ \cite{Freedman:2000cf}, one should keep in mind that the measurements made to determine this value are all made at less than a few hundred megapasec ($z\approx 0.01$), whereas the size of the void in this model is about $1 \, {\rm Gpc}$ ($z\approx 0.02$). The Hubble parameter inside the void in this model turns out to be $65 \, {\rm km \, s^{-1} \, Mpc^{-1}}$, which is consistent with the Hubble Key project value.

The authors of \cite{Quercellini:2008ty} numerically compute the cosmic parallax for two different models. They find an angle-dependent signal with maximal value roughly $0.18 \, {\rm \mu as}$ for sources at redshift of $1$. It is claimed that this signal is within reach of the forthcoming Gaia mission, provided Gaia produces a sky map of the expected number of quasars (roughly $500,000$) with final positional error less than $30 \, {\rm \mu as}$, and that such measurements are repeated after $10$ years.


\section{Bianchi Type I Models}
\label{bianchi}

As we have discussed, anisotropic expansion of the universe around a given observer contributes to cosmic parallax. In the case of LTB models, this may allow us to constrain the distance of our galaxy from the center of the void. The further we are from the center of the void, the more anisotropic the universe would look to us. Of course, LTB spacetimes are not the only ones that can give rise to anisotropic expansion around a point. This raises the question of how we might interpret any observation of cosmic parallax that cannot be accounted for by peculiar velocities alone, since such an observation would not itself be evidence that we live in an LTB universe. To understand this therefore, we study cosmic parallax in alternative anisotropic cosmologies and compare our results with those obtained in~\cite{Quercellini:2008ty}.

For simplicity we consider anisotropic spacetimes that are spatially homogeneous~\cite{Ellis:1968vb, Landau}. In particular, we focus on a subclass of Bianchi Type I spacetimes, the metric for which may be written as 
\be
   ds^2 = -dt^2 + a_1^2(t) dx^2 + a_2^2(t) dy^2 + a_3^2(t) dz^2 \ .
   \label{bianchimetric}
\ee
In general, $a_1(t)$, $a_2(t)$, and $a_3(t)$ are independent scale factors, describing how the three spatial directions scale with time, which reduces to the standard FRW case when $a_1(t)=a_2(t)=a_3(t)$. We specialize to the case when only one of the scale factors differs from the others, say $a_1(t)=a_2(t)\neq a_3(t)$, in which case the expansion is axisymmetric. We do so because we want to compare our results to observations made by an off-center observer in an LTB void. Such an observer will experience axisymmetric cosmic expansion, and so the most direct comparison will be to an axisymmetric Bianchi-I universe. There is, however, an important difference in the symmetries around observers in these two spacetimes. In the axisymmetric Bianchi-I universe there is an additional plane of symmetry normal to the axis of symmetry. The same is not true for an off-center observer in an LTB universe. To see this, it is sufficient to consider the extremal case of an observer outside the void, who can obviously distinguish the two directions along the polar axis: toward the void and away from it. This suggests that cosmic parallax in these two types of anisotropic models will differ at least qualitatively, if not in magnitude.

Setting $a_1=a_2=a(t)$ and $a_3=b(t)$ in (\ref{bianchimetric}), the Einstein equations become
\bea
   H_a^2 &+& 2 H_a H_b = 8\pi G \rho
   \label{einstein1}
\\
   2\dot{H_a} &+& 3H_a^2 = -8\pi G P_z
   \label{einstein2}
\\
   \dot{H_a} &+& H_a^2 + \dot{H_b} + H_b^2 + H_a H_b = -8\pi G P_x\ ,
   \label{einstein3}
\eea
where an overdot denotes a derivative with respect to $t$, $P_x=P_y$ and $P_z$ are anisotropic pressures in the different directions, and we have defined the Hubble parameters $H_a\equiv\dot{a}/a$ and $H_b\equiv\dot{b}/b$. The conservation of energy equation in this case is
\be
   \dot\rho = -2H_a(\rho + P_x) - H_b(\rho + P_z)\ .
   \label{conservation}
\ee
The observational success of FRW cosmology places tight constraints on how anisotropic the universe can be. In order to restrict ourselves to solutions that remain close to an FRW cosmology, we split each of these exact equations into an FRW part which evolves according the FRW equations of motion, and a non-FRW part which we require to remain small, in a sense that we will now make clear. A similar approach was used in~\cite{Barrow:1997sy}. We define
\bea
   H_a(t) &=& \bar{H}(t) + \epsilon f(t)
\\
   H_b(t) &=& \bar{H}(t) + \epsilon g(t)
\\
   \rho(t) &=& \bar{\rho}(t) + \epsilon r(t)
\\
   P_x(t) &=& P_y(t) = \bar{P}(t)
\\
   P_z(t) &=& \bar{P}(t) + \epsilon s(t)
\eea
where overbars denote the FRW quantities and $\epsilon$ is a small perturbative parameter for which we will determine an upper bound later. Substituting these definitions into equations (\ref{einstein1})-(\ref{conservation}) and collecting powers of $\epsilon$ gives the zeroth-order (or background) equations, which are just the usual ones of the FRW metric, and the first-order equations
\bea
   2\bar{H}(f + 2g) = 8\pi G r
   \label{pert1}
\\
   6\bar{H}f + 2\dot{f} = -8\pi G s
   \label{pert2}
\\
   3\bar{H}(f + g) + \dot{f} + \dot{g} = 0
   \label{pert3}
\\
   \dot{r} = -3\bar{H} r - \bar{\rho}(2f + g) - \bar{H} s
   \label{pert4}
\eea
These constitute four equations in four variables, but only three of these equations are independent.  To close the system we need additional information, which we obtain by assuming an equation of state of the form
\be
   s(t) = \sigma r(t)
   \label{eos}
\ee
where the parameter $\sigma$ is taken to be constant. Note that this is analogous to $\bar{P}=w\bar{\rho}$, except that it relates the anisotropic component of the pressure to the non-FRW correction of the energy density. The value of $\sigma$ will be important in determining whether the non-FRW parts of equations (\ref{pert1})-(\ref{pert4}) grow or decay in time. 

Realistic models will be those for which the amount of anisotropy is sufficiently small in the past and present. Assuming the anisotropy is set (for example by inflation) to be sufficiently small at some early epoch, the question then is whether the anisotropy grows or not. In our set-up this corresponds to asking whether the non-FRW parts of equations (\ref{pert1})-(\ref{pert4}) grow or not, and if so, how quickly. We will see that $\sigma$ governs the general behavior of the non-FRW quantities, but for a given $\sigma$, the details will depend on the background (FRW) solution. Since we are integrating from the present up to redshifts of order $1$, our background is well described by the $\Lambda$CDM model. Using this background we can analytically find the asymptotic behavior of the non-FRW quantities for different values of the equation state parameter $\sigma$. 

The background energy density and pressure for $\Lambda$CDM are 
\bea
   \bar{\rho} &=& \bar{\rho}_m + \rho_{\Lambda}  \nonumber
\\
   \bar{P} &=& -\rho_{\Lambda}\ ,  \nonumber
\eea
where $\bar{\rho}_m$ is the background energy density of matter and $\rho_{\Lambda}$ is the effective energy density of the cosmological constant. The background equations become
\bea
   3\bar{H}^2 = 8\pi G (\rho_{\Lambda} + \bar{\rho}_m)
   \label{backgd1}
\\
   2\dot{\bar{H}} + 3\bar{H}^2 = 8\pi G \rho_{\Lambda}
   \label{backgd2}
\\
   \dot{\bar{\rho}}_m = -3\bar{H}\bar{\rho}_m\ .
   \label{backgd3}
\eea
The solutions can be written simply
\bea
   \bar{H}(t) &=& A\coth{\left(\frac{3}{2}At\right)}
\\
   \bar{\rho}_m &=& \rho_{\Lambda}\left[\sinh{\left(\frac{3}{2}At\right)}\right]^{-2}
\\
   A &=& \sqrt{\frac{8\pi G\rho_{\Lambda}}{3}}\ .
\eea
Using (\ref{eos}) and (\ref{pert1}) in (\ref{pert4}) then gives
\be
   \dot{r} = -\left[(3+\sigma)A \coth{\left(\frac{3}{2}At\right)} 
   +\frac{12A\pi G}{\sinh{\left(\frac{3}{2}At\right)}}\right] r(t)\ ,
\ee
which can be integrated to find 
\be
   r(t) = c_1 \cosh{\left(\frac{3}{2}At\right)} 
   \left[\sinh{\left(\frac{3}{2}At\right)}\right]^{-(3+\frac{2}{3}\sigma)}\ .
   \label{soln1}
\ee
Using the equation of state (\ref{eos}) immeditely gives $s(t)$. Equations (\ref{pert2}) and (\ref{pert3}) can then be integrated to find the remaining solutions (for $\sigma\neq-9/2$)
\bea
   f(t) &=& \frac{A}{\left[\sinh{\left(\frac{3}{2}At\right)}\right]^{2}}
   \left(c_2 + \frac{3 c_1}{2 \left[\sinh{\left(\frac{3}{2}At\right)}\right]^{\frac{2}{3}\sigma}}\right)
   \label{soln2}
\\
   g(t) &=& -f(t) - \frac{A c_2}{\left[\sinh{\left(\frac{3}{2}At\right)}\right]^{2}}\ .
   \label{soln3}
\eea
Now we consider the asymptotic behavior of these solutions for $t\rightarrow\infty$ and $t\rightarrow 0$ in turn.

The behavior as $t\rightarrow\infty$ is
\bea
   r(t) &\sim& s(t) 
   \sim c_1\left[\sinh{\left(\frac{3}{2}At\right)}\right]^{-2(1+\frac{1}{3}\sigma)}\ .
\\
   f(t) &\sim& -g(t) \sim c_1\left[\sinh{\left(\frac{3}{2}At\right)}\right]^{\textrm{Max}\left[-2,-2(1+\frac{1}{3}\sigma)\right]}\ .
\eea
Here we see that $\sigma=-3$ is the boundary between growing and decaying solutions. If we require all of the non-FRW quantities to decay as $t\rightarrow\infty$, then we must restrict ourselves to $\sigma>-3$. 

As $t\rightarrow 0$, the situation is slightly more complicated. The functions $f(t)$ and $g(t)$ both contain a piece that depends on the value of $\sigma$ and a piece that does not. The part that does not behaves as
\be
   c_2 \frac{A}{\left[\sinh{\left(\frac{3}{2}At\right)}\right]^{2}} 
   \sim c_2 t^{-2}\ , 
   \,\,\,\,\,\, t \rightarrow 0\ ,  \nonumber
\ee
whereas the part that depends on $\sigma$ behaves as
\be
   c_1 \frac{3A}{2\left[\sinh{\left(\frac{3}{2}At\right)}\right]^{2(1+\frac{1}{3}\sigma)}} 
   \sim c_1 t^{-2(1+\frac{1}{3}\sigma)}\ , 
   \,\,\,\,\,\, t \rightarrow 0\ .  \nonumber
\ee
We require that the expansion history is close to FRW in the far past, which amounts to demanding that $|f/\bar{H}|$, $|g/\bar{H}|$, etc remain $\lesssim\mathcal{O}(1)$ as $t\rightarrow 0$. From the first term, this requires that $c_2=0$, while the second term requires that $\sigma<-3/2$. Therefore solutions with decaying anisotropy have an equation of state parameter lying in the range $-3<\sigma<-3/2$. Turning to $r(t)$ and $s(t)$, as $t\rightarrow 0$ we have  
\be
   r(t) \sim s(t) \sim c_1\left[\sinh{\left(\frac{3}{2}At\right)}\right]^{-(3+\frac{2}{3}\sigma)}\ .
\ee
When $\sigma>-3$ these solutions diverge as $t\rightarrow 0$, but if $\sigma<-3/2$ they diverge slower than $\bar{\rho}$ and $\bar{P}$, respectively. So the condition
\be
   -3 < \sigma < -\frac{3}{2}
\ee
will ensure that all non-FRW quantities remain small in the far past, as required. We will restrict ourselves to these models in the rest of the paper. 

What remains is to fix $c_1$ and the constant $A$ in the solutions (\ref{soln1})-(\ref{soln3}). We do this by imposing observational constraints on the models of interest. First we impose a condition at the surface of last scattering. For an anisotropically expanding universe to be viable it must at the very least predict angular variations in the temperature of the CMB no bigger than $10^{-5}$. We can estimate the maximum temperature difference at the time of last scattering by
\be
   \Delta T_0 = \left| T^{xy}_0 - T^{z}_0 \right|
   = \left| T_{\rm lss} \left(\frac{a_{\rm lss}}{a_0}\right) 
   - T_{\rm lss} \left(\frac{b_{\rm lss}}{b_0}\right) \right| \ ,
\ee
where $T^{z}$ is the temperature along the axis of symmetry and $T^{xy}$ is the temperature in the transverse plane; subscripts ``0'' and ``lss'' refer to quantities today and at the last scattering surface, respectively. Recall that $b(t)$ is the scale factor along the axis of symmetry and $a(t)$ is the scale factor in the transverse plane. From the definitions of $H_a$ and $H_b$, we have
\bea
   a(t) &=& a_0\exp{\int_{t_0}^{t} (\bar{H}(t') + \epsilon f(t')) dt'}
\\
   b(t) &=& b_0\exp{\int_{t_0}^{t} (\bar{H}(t') + \epsilon g(t')) dt'}\ .
\eea
Using these, and choosing $a_0=b_0=1$, we can re-write the expression for $\Delta T_0$ to first order in $\epsilon$ as
\bea
   \Delta T_0 &=& T_{\rm lss} \left( \frac{b_{\rm lss}}{b_0} \right) 
   \left| 1 - \left( \frac{a_{\rm lss}}{b_{\rm lss}} 
   \frac{b_0}{a_0} \right) \right| 
   \simeq 2\epsilon T^{z}_0\left|\int_{t_0}^{t_{\rm lss}} f(t) dt\right|\ ,
\eea
where we have used the fact that $g(t)=-f(t)$, which is a consequence of requiring $c_2=0$. Inserting the solution for $f(t)$ gives
\be
   \frac{\Delta T_0}{T^{z}_0}
   \simeq 3A \epsilon c_1\int_{t_0}^{t_{\rm lss}}\left[\sinh{\left(\frac{3}{2}At\right)}\right]^{-2(1+\frac{\sigma}{3})}dt\ .
   \label{condition1}
\ee
Demanding the left-hand side to be at most $1.3 \times 10^{-6}$ \cite{Hinshaw:2008kr} yields one condition on the product $\epsilon c_1$ and $A$ for a given equation of state parameter $\sigma$. 

In order to break the degeneracy between $\epsilon c_1$ and $A$, we need one further condition. Equation (\ref{condition1}) already requires the difference between $H_a$ and $H_b$ to be small -- well within the accepted uncertainty in the measured value of the Hubble parameter~\cite{Freedman:2000cf}. We find it convenient to choose to set the arithmetic average of the Hubble parameters in the three directions equal to the observed value. 
\be
   H_{\rm obs} = \frac{2H_a + H_b}{3} 
   = \bar{H} + \frac{\epsilon}{3}f(t)\ ,
   \label{avg}
\ee
where again we have used the fact that when $c_2=0$, $g(t)=-f(t)$. Alternative choices, such as setting $H_a$ or $H_b$ equal to the measured value of the Hubble parameter, would not change the order of magnitude of $\epsilon c_1$ and $A$ and would leave the final result essentially unaltered. Inserting the solution for $f(t)$ into equation (\ref{avg}) gives a second condition on $\epsilon c_1$ and $A$
\be
   \epsilon c_1 = \frac{2}{A}(H_{\rm obs} - \bar{H})
   \left[\sinh{\left(\frac{3}{2}At\right)}\right]^{2(1+\frac{\sigma}{3})}\ .
   \label{condition2}
\ee
We can (numerically) solve this equation for $A$ in terms of $\epsilon c_1$ and then substitute it back into the first condition (\ref{condition1}) to obtain an upper bound on $\epsilon c_1$. By taking the maximal allowed value for $\epsilon c_1$, we can then find $A$ using (\ref{condition2}). For $\sigma=-2$ we find $A\simeq62$ and $\epsilon c_1\simeq1.3\times10^{-6}$. We always take $c_1$ to be $\mathcal{O}(1)$, and so in this case we choose $A=1.3$ and $\epsilon=10^{-6}$. 

In this way we can fully determine solutions to the non-FRW quantities for a given equation of state parameter $\sigma$. By restricting ourselves to $-3<\sigma<-3/2$ we have chosen to focus on models for which these solutions decay in the distant future and which diverge slower than the respective background quantities in the far past as one approaches the initial sigularity. These models seem to be the most conservative realizations of a Bianchi-I cosmology, in the sense that they are the easiest to make consistent with observations, or alternatively, the most difficult to rule out. One might try to push the boundaries slightly, for example by considering models with anisotropies that approach a constant in the distant future rather than vanishing. It may be that such models can be carefully tuned to match observations. We do not consider these more general models here, since our main interest is not model-building, but rather to explore a general effect (cosmic parallax) arising in a Bianchi-I universe.    


\section{Geodesics and Cosmic Parallax in Axisymmetric Bianchi-I Models}
\label{parallax}

As in section \ref{ltb}, in order to analyze cosmic parallax we need to find null geodesics in the spacetime that connect an observer and various sources at two different times. Letting latin indices run over ${1,2,3}$ and $(x^0,x^1,x^2,x^3)=(t,x,y,z)$, the non-zero Christoffel symbol components for the Bianchi-I metric in (\ref{bianchimetric}) are
\be
   \Gamma^0{}_{ij} = a_i^2 H_i\delta_{ij}\ , \quad 
   \Gamma^i{}_{0j} = H_i\delta^i{}_j \ , \nonumber
\ee
where no sum on the index $i$ is implied and $H_i$ is defined as above. The four geodesic equations are then
\bea
   \frac{d^2 t}{d\lambda^2} &=& -\sum_i H_i\left(a_i\frac{dx^i}{d\lambda}\right)^2
   \label{geo 1}
\\
   \frac{d^2 x^i}{d\lambda^2} &=& -2 H_i \frac{dt}{d\lambda}\frac{dx^i}{d\lambda}\ ,
   \label{geo 2}
\eea
with the null geodesic condition, $u^\mu u_\mu=0$, becoming
\be
   \left(\frac{dt}{d\lambda}\right)^2 = \sum_i \left(a_i\frac{dx^i}{d\lambda}\right)^2\ .
   \label{constraint}
\ee
As before, $u^\mu\equiv dx^\mu/d\lambda$. We specialize to the axisymmetric case by setting $a_1=a_2=a(t)$ and $a_3=b(t)$, and also $H_1=H_2=H_a$ and $H_3=H_b$. The scale factors, $a(t)$ and $b(t)$, and Hubble parameters, $H_a(t)$ and $H_b(t)$, are fixed after choosing $\sigma$ and solving the full set of equations as in the previous section.

After fixing the background Bianchi-I spacetime, equations (\ref{geo 1}) and (\ref{geo 2}) yield four second-order differential equations and one constraint equation in four dependent variables. To solve this system we must in principle specify initial conditions for the four dependent variables as well as initial velocities (derivatives with respect to $\lambda$) giving a total of eight conditions. However, using the constraint equation the system can be reduced to seven independent first-order equations.

Considering the $u^\mu (t)\equiv dx^{\mu} (t) / d\lambda$ as functions of time along the geodesic, equations~(\ref{geo 2}) can be integrated immediately to give
\bea
   u^i(t) &=& u^i_0 \, e^{-2\int_{t_0}^t H_i(t^\prime)dt^\prime}  \nonumber
\\
   &=& u^i_0 \, a_i^{-2}(t) \ ,
   \label{eq1}
\eea
which can then be used in equation (\ref{constraint}) to give
\be
   \left(\frac{dt}{d\lambda}\right)^2 = 
   \sum_i \left(\frac{u^i_0}{a_i(t(\lambda))}\right)^2  \ .
   \label{eq0}
\ee
As in the LTB case it is again useful to find an expression for the redshift as it will be one of our observational inputs. To find this expression, as usual, we consider two photons emitted from a source at times $t$ and $t+\tau$, respectively, where $\tau$ is taken to be infinitesmally small. The trajectory of the first photon is described by equation (\ref{eq0}), while to first order in $\tau$, the trajectory of the second photon is described by the geodesic equation 
\be
   \left(\frac{dt}{d\lambda}\right)^2 + 2\frac{dt}{d\lambda}\frac{d\tau}{d\lambda} 
   = \sum_i \left(u^i a_i\right)^2 \left(1+2\tau H_i \right)\ .
   \label{redgeo}
\ee
Here the variation in time corresponds to a time-delay, and so to a change in geodesic, and not to a change of the time coordinate along a fixed geodesic. Since the $u^i(\lambda)$ are directional derivates along a given geodesic, they remain unaffected by time variations (derivatives) in obtainig this equation. Using the standard definition of redshift, $1+z(\lambda_{\rm em})\equiv \tau(\lambda_{\rm ob})/\tau(\lambda_{\rm em})$, we find the relation
\be
   \frac{d \log(1+z(\lambda_{\rm em}))}{d\lambda_{\rm em}} 
   = -\frac{1}{\tau(\lambda_{\rm em})} \frac{d\tau(\lambda_{\rm em})}{d\lambda_{\rm em}}\ ,
\ee
where $\lambda_{\rm em}$ and $\lambda_{\rm ob}$ are the values of the affine parameter at the emission and observation event, respectively. 
Using this in (\ref{redgeo}) gives
\bea
   \frac{d\log(1+z)}{d\lambda} 
   &=& -\frac{ \sum_i  \left(u^i a_i\right)^2 H_i}{\sqrt{\sum_i  \left(u^i a_i\right)^2}}  \nonumber
\\
   &=& \frac{d\log{\sqrt{\sum_i  \left(u^i a_i\right)^2}}}{d\lambda}\ .
   \label{redeq}
\eea
In the last step we have also made use of (\ref{eq1}) and the chain rule
\be
   \frac{d}{d\lambda}(u^i(\lambda) a_i(\lambda))
   = - \frac{dt}{d\lambda}(u^i(t) a_i(t)H_i(t)) \ .  \nonumber
\ee
With the initial condition $z(\lambda_0)=0$, integrating~(\ref{redeq}) then gives
\be
   (1 + z) = \frac{\sqrt{\sum_i  \left(u^i a_i\right)^2}}{\sqrt{\sum_i  \left(u^i_0\right)^2}}\ .
   \label{eq2}
\ee

As in section \ref{ltb}, we would like to express our results not only in terms of redshift but also in terms of angles, since these are the actual observables. In general there are four pieces of data for each object in the sky, namely the time of observation, two angles with respect to an arbitrary coordinate system, and the observed redshift of the source. In the case of cylindrical symmetry one angle is enough. In contrast with LTB spacetimes, not much is gained in this case by rewriting the geodesic equations in terms of angles and redshift. Instead, we numerically integrate the equations in the above coordinates and then express the results in terms of angles and redshift.

Our procedure for computing the cosmic parallax for these models is analogous to that in section \ref{ltb}. We work in local Cartesian coordinates $(t,x_1,x_2,x_3)$ in which the observer is located at the origin. As seen above, the system of four second-order differential equations plus a constraint reduces to seven independent first-order equations, some of which can be integrated immediately by hand. We are then left with the problem of fixing initial conditions for our complete set of equations
\bea
   \frac{d t}{d\lambda} &=&- \sqrt{\sum_i \left(u^i a^i \right)^2} 
\\ 
   u^i(\lambda) &=& \frac{u^i_0}{a_i^2}
\\
   \frac{d x^i}{d\lambda} &=& u^i  \ ,
\eea
where the subscripts ${}_0$ refer to quantities at the initial observation event, corresponding to $\lambda=0$. To close the system requires seven initial conditions that unfortunately cannot be all specified at the observation event. By construction we have $x^i{}_0=0$, leaving four remaining conditions, three of which are obtained by specifying $t_0$ and two initial spatial velocities in terms of measurable quantities (angles), while the last one is given by the observed redshift.  We integrate backwards along the initial geodesic until reaching the desired redshift and then find the comoving coordinates of the source. To find the final geodesic that connects the same source with the observer at a later time $t_0+\Delta t$, we solve the corresponding boundary-value problem (namely, to find solutions of the null geodesic equations with two fixed endpoints). We then find the velocities along the final geodesics at the time of observation.

\begin{figure}[h]
   \begin{center}
   \includegraphics[width=0.4\textwidth]{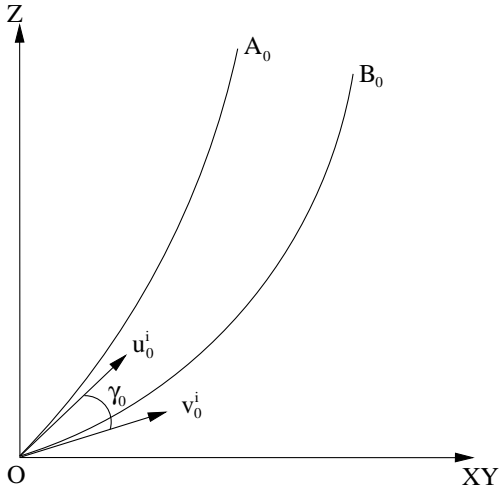}
   \caption{Initial angle $\gamma _0$ defined by the velocity vectors at observing time for two sources}
   \label{kas1}
   \end{center}
\end{figure}

\begin{figure}[h]
   \begin{center}
   \includegraphics[width=0.4\textwidth]{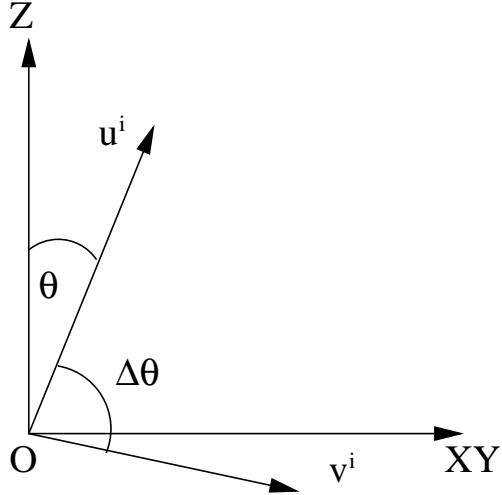}
   \caption{Location of sources as seen by an observer at the measuring event. The vertical axis points along the axis of symmetry. We are considering sources at equal redshifts of $z=1$ and in a plane defined by a fixed value of the polar angle $\phi$.}
   \label{kas2}
   \end{center}
\end{figure}

\begin{figure}[h]
   \begin{center}
   \includegraphics[scale=0.8]{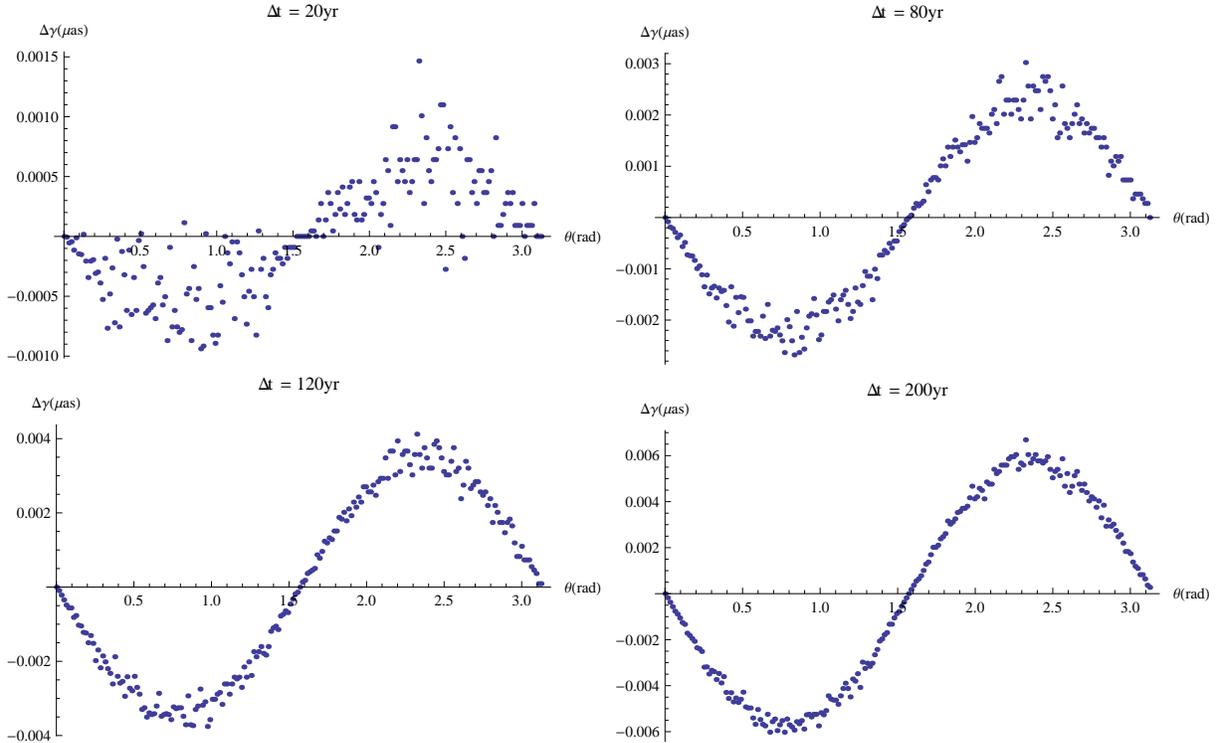}
   \caption{A sequence of cosmic parallax signals for different values of $\Delta{t}$. Top row, left to right: $\Delta{t}=20$yrs, $\Delta{t}=80$yrs. Bottom row, left to right: $\Delta{t}=120$yrs, $\Delta{t}=200$yrs. The signal-to-noise ratio becomes smaller as $\Delta{t}$ becomes smaller.}
   \label{array}
   \end{center}
\end{figure}

\begin{figure}[h]
   \begin{center}
   \includegraphics[scale=0.6]{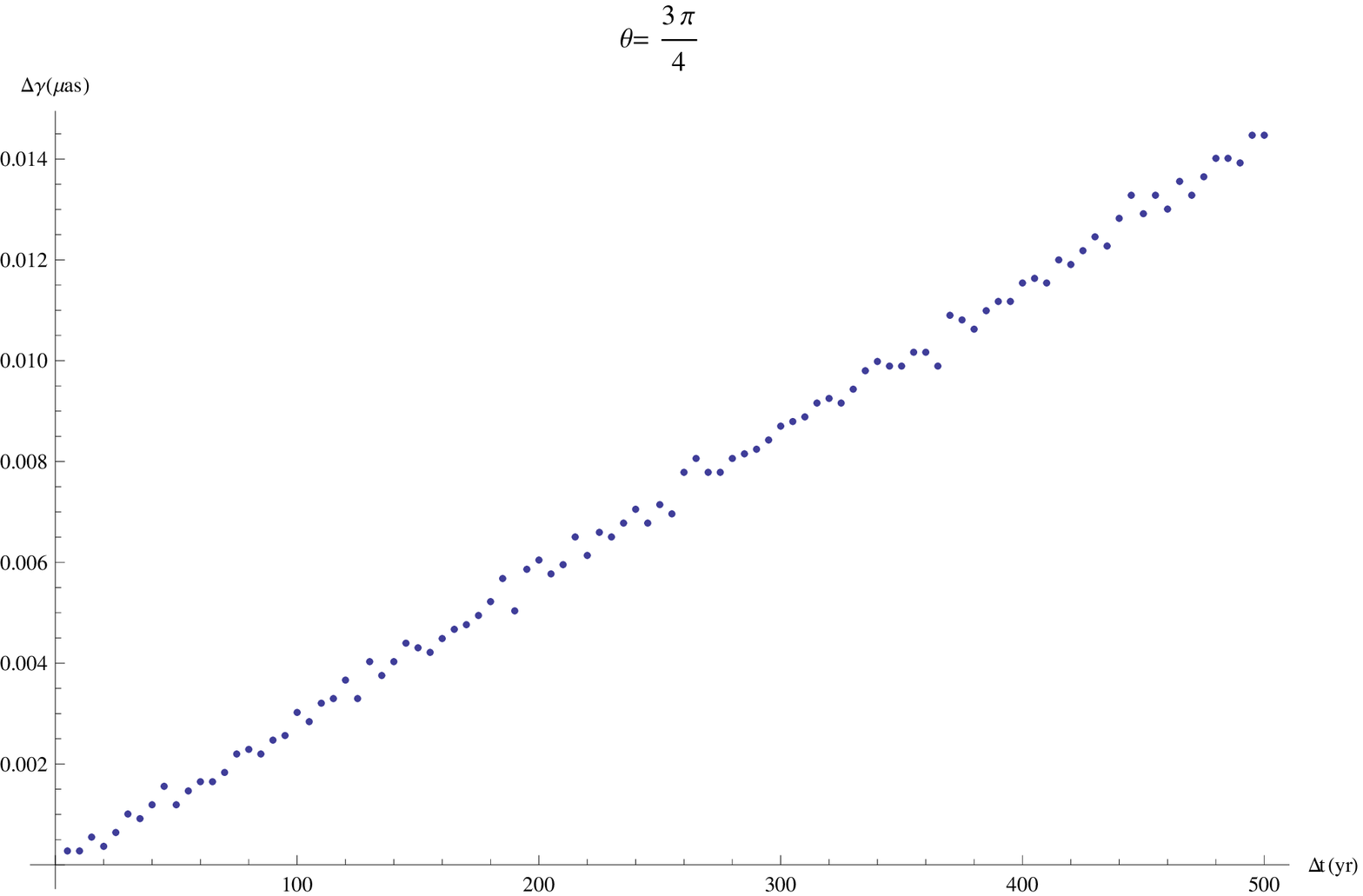}
   \caption{Cosmic parallax as a function of $\Delta{t}$. Here $\epsilon=10^{-6}$ and sources are at $z=1$, separation angle ($\Delta\theta$) of $90$ degrees.}
   \label{slice}
   \end{center}
\end{figure}

This procedure is repeated for two sources, yielding four sets of velocities: one set for each initial geodesic and one set for each final geodesic. The velocities $u^i$ and $v^i$ along two geodesics at the same observing time (see figure~\ref{kas1}) are related to the angle $\gamma$ between them by
\be
   \cos{(\gamma)} = 
   \frac{\sum_i a^2_i u^i v^i}{\sqrt{\left(\sum_j (a_j u^j)^2\right)\Bigl(\sum_k (a_k v^k)^2\Bigr)}} \ .
\ee
We then calculate the cosmic parallax $\Delta\gamma$, as defined in (\ref{cpdef}), by taking the difference of the angle between the two sources at the two different times. Finally, we plot the cosmic parallax as a function of the polar coordinate $\theta$ for one of the sources. By convention we choose the second, or trailing, source as our reference, as shown in figure~\ref{kas2}.
We find the parallax for two hypothetical sources with the same redshift, initially separated by a given angle on the sky, and for a given $\Delta{t}$ and $\epsilon$.  In figure~\ref{array} we plot the parallax for two hypothetical sources at $z=1$, with an initial separation of $90$ degrees on the sky, for $\epsilon=10^{-6}$ and various values of $\Delta{t}$.
For $\Delta{t}=500$ years and $\epsilon=10^{-6}$ we find this maximal value to be of the order of $6 \times 10^{-14}$ radians. Of course, our true goal is to find the maximal signal for reasonable time scales, say $\Delta{t} \sim 10$ years, but unfortunately numerical noise dominates over the signal for $\Delta{t}$ of that magnitude. This problem is shown in figure~\ref{array} where the results from our code for the cosmic parallax are plotted for different values of $\Delta{t}$.
One can see that by $\Delta{t}=20 \, {\rm years}$ the signal-to-noise ratio becomes very poor, making it difficult to extract trustworthy predictions. Therefore, although we can directly compute the signal for this model at $\Delta{t}=10 \, {\rm years}$, we prefer to calculate the cosmic parallax for several values of $\Delta{t}$ between $5$ and $500$ years and interpolate the value at $10$ years. Since our primary goal is to find an order of magnitude estimate for the effect, this approach should be acceptable. Figure~\ref{slice} shows the values of the cosmic parallax between two sources for decreasing values of $\Delta{t}$ (keeping all other values fixed). 
For each value of $\theta$, we use a linear fit passing through the origin (because cosmic parallax must vanish for $\Delta{t}=0$) to find the interpolated value at $10$ years. We could just as well perform an extrapolation by omitting data below some cut-off, say $\Delta{t}=50 \, {\rm years}$. This does not have any noticeable effects on our results. 

\begin{figure}[h]
   \begin{center}
   \includegraphics[scale=0.6]{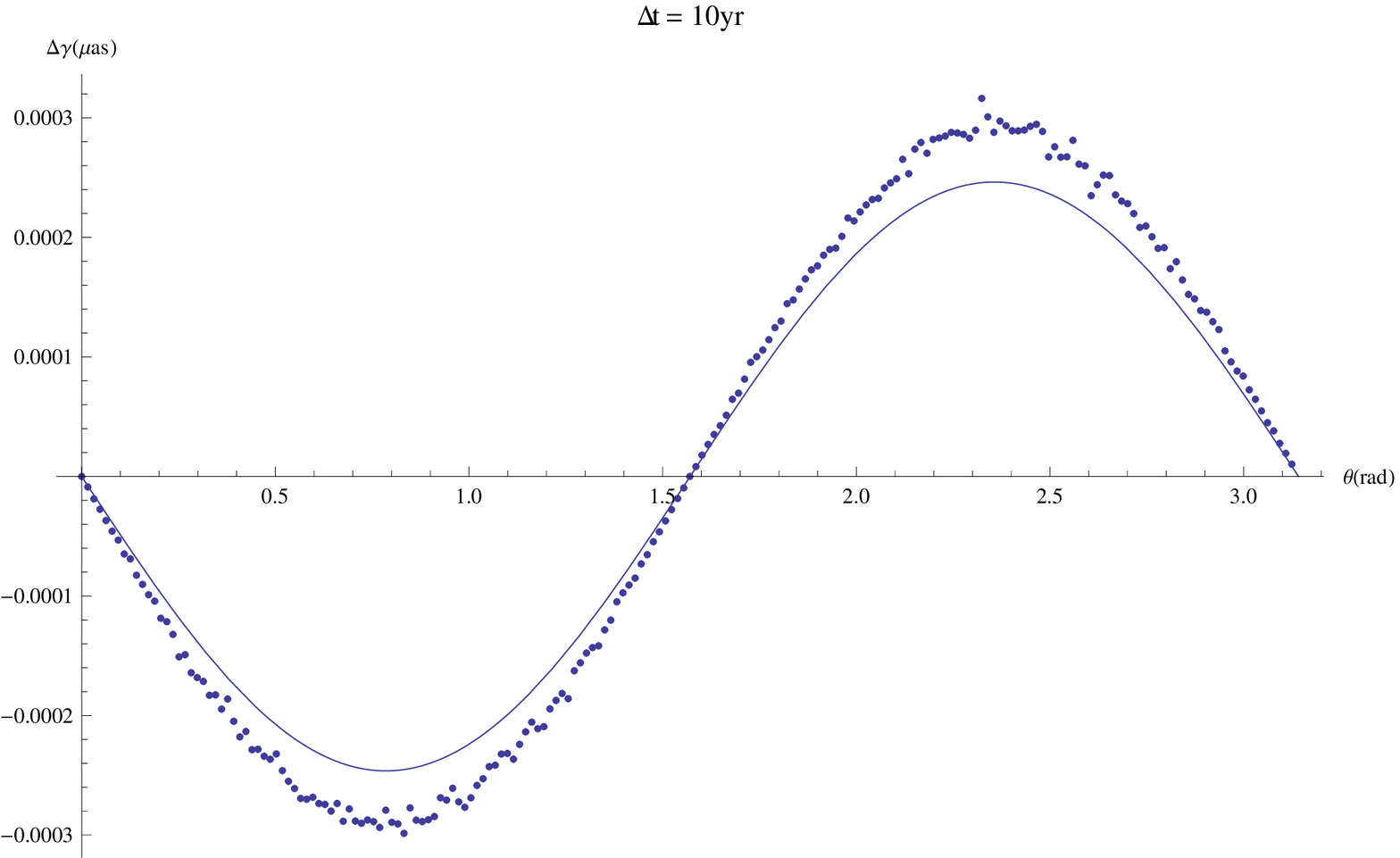}
   \caption{Estimate of the cosmic parallax for $\Delta{t}=10$ years, $\epsilon=10^{-6}$, and for sources at $z=1$ with a separation angle of $90$ degrees, plotted as a function of the angle between the trailing source and the symmetry axis (i.e., the angle between the leading source and the symmetry axis is advanced by $90$ degrees). Each of the plotted points is an interpolated value using $100$ trials at values of $\Delta{t}$ spread from $5-500$ years.  The solid line is a fit of a sine function through the raw data for the cosmic parallax calculated for $\Delta{t}=10$ years.}
   \label{extrap}
   \end{center}
\end{figure}

After repeating the interpolation for all of our data points, the result is plotted in figure~\ref{extrap}. In the same figure we also fit a sine function through the parallax data obtained directly from our code for $\Delta{t}=10 \, {\rm years}$. This fit is quite close to the interpolated values, suggesting that the numerical noise is randomly distributed about zero and strengthening our confidence on the correctness of the linear interpolation procedure.

For the largest allowed values of $\epsilon$ consistent with the observed anisotropy of the CMB (i.e., $\epsilon\sim10^{-6}$) and for time delays on the order of $10$ years, we find that the maximal cosmic parallax is on the order of $10^{-15}$ radians, or equivalently $10^{-4}$ microarcseconds. This is three orders of magnitude smaller than the maximal cosmic parallax seen by \cite{Quercellini:2008ty} for LTB models. It is also three orders of magnitude smaller than the expected level of cosmic parallax from peculiar velocities alone in a $\Lambda$CDM universe \cite{Ding:2009xs}. In other words, the contribution to cosmic parallax in this model due to anisotropic expansion is sub-dominant to the contribution from peculiar velocities \footnote{Here we assume that the Bianchi-I models we consider have roughly the same peculiar velocity-redshift relation as an FRW universe}.

The qualitative behavior of the cosmic parallax in this model is also quite different from that of LTB models as put forth in \cite{Quercellini:2008ty}. Both LTB and Bianchi models show a sinusoidal (or at least quasi-sinusoidal) cosmic parallax. However, whereas LTB models exhibit a $2\pi$-periodic behavior, here we see that Bianchi models exhibit a $\pi$-periodic behavior. This is to be expected due to the symmetries of the two types of spacetimes, as alluded to earlier in the paper. The LTB spacetime is axisymmetric about an off-center observer but is not plane-symmetric about the plane normal to the symmetry axis. If we align the z-axis along the symmetry axis, then in spherical-polar coordinates this amounts to saying that the spacetime is invariant under changes in $\phi$ (the azimuthal angle) but has no symmetries under (nontrivial) changes in $\theta$ (the polar angle). In other words, one would expect cosmic expansion to be $2\pi$-periodic in $\theta$, which is just what we see for the cosmic parallax in these models. The Bianchi spacetimes, on the other hand, are both axisymmetric and plane-symmetric about the plane normal to the axis of symmetry. So one would expect cosmic expansion to be $\pi$-periodic in $\theta$, which is what we see for the cosmic parallax in these models.

Although here we have only considered a particular model, preliminary investigations into other models suggest that these results are robust. For example, one might consider models with other constant values for the equation of state parameter $\sigma$, a time-varying rather than constant $\sigma$, or an equation of state that takes a different form than equation (\ref{eos}). None of these modifications seem to affect the order of magnitude of the cosmic parallax (which, due to the symmetry of the metric, is the only free parameter). What this suggests is that the contribution to the cosmic parallax from viable Bianchi-I models is much smaller than the contribution from viable LTB-void models. If the observed cosmic parallax deviates from what is expected in an FRW universe, it is unlikely that this is due to our living in a Bianchi-I spacetime.


\section{Conclusions}

In order for the standard FRW cosmology to agree with observations, the expansion of the universe must be accelerating. So far, all suggested mechanisms to drive such acceleration involve new physics; either the existence of exotic new components of the cosmic energy budget, modifications to Einstein's theory of gravity, or a cosmological constant.  Alternatively, it has been suggested that the phenomenon of cosmic acceleration is due to interpreting results in the FRW model, when in fact the correct underlying cosmic geometry could be that of a void model, such as that idealized by an LTB spacetime. These models are relatively simple and make a host of predictions that can be used to either test or constrain them. One prediction that may soon be testable is cosmic parallax. For observers located $10$ Mpc from the center of a $1$ Gpc LTB void, this effect would have a magnitude of the order of $10^{-1}\mu$as per decade for sources at redshift 1.  Since the effect in an FRW universe (due to peculiar velocities) is expected to be roughly the same order of magnitude, one might hope to subtract the signal due to peculiar velocities from the signal due to anisotropic expansion about a point. Measuring this additional contribution to the cosmic parallax would clearly indicate a departure from FRW. 

However, what is less clear is how we might interpret detection of an additional contribution to cosmic parallax. Here we have examined an axisymmetric Bianchi-I universe as an alternative explanation for any cosmic parallax component that is not attributable to peculiar velocities. We find that for a class of models whose anisotropy is consistent with the observed temperature anisotropies of the CMB and Hubble expansion today, the maximum amount of cosmic parallax is three orders of magnitue smaller than the maximal signal in LTB models. Perhaps more importantly, the maximum effect is also three orders of magnitude smaller than the expected level of cosmic parallax from peculiar velocities in an FRW universe. Although we have focused our discussion in this paper on a particular model, we have found these results to be fairly model-independent. Thus it seems unlikely that measurements of cosmic parallax can constrain Bianchi-I models that are not already ruled out by the CMB or other cosmological observations. Therefore, while cosmic parallax will be nonzero for any anisotropic expansion, the magnitude of the effect suggested in~\cite{Quercellini:2008ty} appears to be significantly larger, and qualitatively different than in the class of models considered here.


\acknowledgments

We would like to thank Cristian Armendariz-Picon and Miguel Quartin for useful discussions. This work was supported by National Science Foundation grants PHY-0653563 and PHY-0930521, by NASA under ATP grant NNX08AH27G, and by the University of Pennsylvania.



\end{document}